\documentstyle[prd,aps,preprint]{revtex}
\begin{document}
\draft


\title{Applying black hole perturbation theory to numerically generated
spacetimes}

\author{Andrew M. Abrahams\cite{ama}}
\address{Department of Physics and Astronomy, University of North Carolina,
Chapel Hill, NC 27599-3255}
\author{Richard H. Price}
\address{Department of Physics, University of Utah, Salt Lake City, Utah,
84112}

\date{August 28, 1995}
\maketitle
\tightenlines

\begin{abstract}
\widetext Nonspherical perturbation theory has been necessary to
understand the meaning of radiation in spacetimes generated through
fully nonlinear numerical relativity. Recently, perturbation
techniques have been found to be successful for the time evolution of
initial data found by nonlinear methods. Anticipating that such an
approach will prove useful in a variety of problems, we give here both
the practical steps, and a discussion of the underlying theory, for
taking numerically generated data on an initial hypersurface as
initial value data and extracting data that can be considered to be
nonspherical perturbations.

\end{abstract}
\section{Introduction}
\label{sec:intro}

The formation of a black hole is, in principle, one of the most
efficient mechanisms for generation of gravitational waves. Such
sources tie together two major research initiatives. Laser
interferometric gravity wave detectors\cite{ligo} hold out a promise
of the detection of gravitational waves from astrophysical events. To
interpret the results of the gravitational wave signals, and to help
find signals in the detector noise, a broad and detailed knowledge
will be needed of astrophysical gravitational waveforms. This is one
of the underlying motivations for the ``grand
challenge''\cite{grandchallenge} in high performance computing, aimed
at computing the coalescence of black hole binaries.

Evolving numerical spacetimes and extracting outgoing radiation
waveforms is indeed a challenge.  In a straightforward numerical
approach, a good estimate of the asymptotic waveform requires long
numerical evolutions so that the emitted waves can be propagated far
from the source.  The necessary long evolutions are difficult for a
number of reasons. General difficulties include throat stretching when
black holes form, numerical instabilities associated with curvilinear
coordinate systems, and the effects of  outer boundary
conditions which are approximate.\cite{ast95}

We suggest here that at least part of the cure for this problem may
lie in the use of the theory and techniques of nonspherical
perturbations of the Schwarzschild spacetime (``NPS''). By this we
mean the techniques for treating spacetimes as deviations, first order
in some smallness parameter, from the Schwarzschild spacetime.  These
techniques differ from ``linearized theory'' which treats
perturbations of the spacetime from Minkowski spacetime and which
cannot describe black holes.  The basic ideas and methods were set
down by many authors and lead to ``wave equations'' for the even
parity\cite{zerilli} and odd parity\cite{rw} perturbations.

NPS has been used to compute outgoing radiation waveforms from a wide
variety of black hole processes, including the scattering of
waves\cite{waves}, particles falling into a hole\cite{particle}, and
stellar collapse to form a hole\cite{stars}.  The general scheme of
NPS also underlies the techniques for extraction of radiation from
numerically evolved spacetimes\cite{ae90}. NPS computations have
recently been used in conjunction with fully numerical evolution, as a
code test\cite{abhss} and as a strong-field radiation extraction
procedure\cite{ast95}.

Here we are interested in another sort of application of NPS
theory. To understand such applications we consider an example: Two
very relativistic neutron stars falling into each other, coalescing
and forming a horizon, as depicted in Fig.~1.
The curve ``hypersurface," in Fig.~1, indicates a spacelike ``initial"
surface.  The spacetime can be divided into three regions by this
initial surface and the horizon. The early evolution, in region I,
below the initial hypersurface, is highly dynamical and
nonspherical. Spherical perturbation theory is clearly
inapplicable. Above the initial surface the spacetime remains highly
nonspherical in region II inside the event horizon, but outside the
event horizon, in region III, it may be justified to consider the
spacetime to be a perturbation of a Schwarzschild spacetime. This is
essentially guaranteed if the initial hypersurface is chosen late
enough, in some sense, after the formation of the horizon.  The
evolution in region III, then, is determined by cauchy data on the
initial hypersurface exterior to the horizon. It is important to note
that this is made possible by the fact that the horizon is a causal
boundary which shields the outer region from the dynamics of the
highly nonspherical central region.

The scheme inherent in this division of spacetime has the potential
greatly to increase the efficiency of the computation of the radiation
generated when strong field sources form black holes. If one starts
from the cauchy data on the initial hypersurface, one can evolve
forward in time with the linear equations of perturbation theory.
Many of the long-time evolution problems of numerical relativity are
avoided and the interpretation of the computed fields in terms of
radiation is immediate.

The approach suggested would then seem to be: Use numerical relativity
up to the initial hypersurface; use the techniques of nonspherical
perturbations in the future of the initial hypersurface. In fact, the
efficiency that can be achieved may be even greater. In the early,
highly nonspherical, pre-initial hypersurface phase of the development
of the spacetime, there may be relatively little generation of
gravitational radiation. By using a computational technique which
suppresses the radiative degrees of freedom one may be able to compute
the early stages of evolution relatively easily.  There are two very
recent examples of just such applications of this viewpoint. Price and
Pullin\cite{price_pullin94} used as initial data the
Misner's\cite{misner} solution to the initial value equations for two
momentarily stationary black holes. Abrahams and
Cook\cite{abrahams_cook94} considered two holes moving towards each
other, and used numerical values of the initial value equations.  In
neither case was there {\em any} use of fully nonlinear numerical
evolution. The rather remarkable success of both computations suggests
that there is something robust about the underlying idea of separating
horizon-forming astrophysical scenarios into an early phase with no
radiation and a late phase with small deviations from sphericity
outside the horizon.  It is plausible that the bulk of the radiation
in most processes is generated only in the very strong-field
interactions around the time of horizon formation and that radiation
generation in the early dynamics can be ignored. One would, however,
think that strong radiation would be emitted during the stages at
which the early horizon is very nonspherical and at which time
nonspherical perturbation theory would seem to be inapplicable. There
should be a tendency for this ``early'' radiation, produced very close
to the horizon, to go inward into the developing black hole, so that
the application of nonspherical perturbation theory to the exterior
really requires that on the initial spacetime the perturbation are
small only well outside the horizon. It would seem that something of
this sort would have to be happening to explain the accuracy of the
Price-Pullin and Abrahams-Cook results.

Whether or not many problems can be treated with {\em no} use of fully
numerical evolution, it appears clear to us that these perturbation
methods will be applied to a variety of problems in which data on the
initial hypersurface is available numerically.  The primary purpose of
this paper is to provide justification and background for earlier work
on this subject and a clear recipe for future applications.  In the
next section we discuss the meaning, and limitations, of extracting a
``perturbation'' from this numerical data and computing radiated
energies.  The explicit process of extracting the perturbations from
the numerical data is given in Sec.~III.  In Sec.~IV we demonstrate
the use of this procedure via application to a specific example, the
Misner initial data.

\section{Initial data as Schwarzschild perturbations}

We outline here the formalism for perturbation theory based on work by
Regge and Wheeler\cite{rw} and by Zerilli\cite{zerilli}, but we will
draw heavily on the gauge invariant reformulation of those earlier
works by Moncrief\cite{moncrief74}.  Our starting point is an initial
hypersurface which can be taken as a surface of constant Schwarzschild
time. We assume that the coordinates ${x^i}$ on that surface are
almost Schwarzschild coordinates $r,\theta,\phi$ and we assume that
the values are known, on this hypersurface and in these coordinates,
for the 3-metric $\gamma_{ij}$ and the extrinsic curvature
$K_{ij}$. The conditions for finding such a hypersurface and such
coordinates will be made explicit in Sec.~III.

Underlying perturbation theory is the idea of a family of metric
functions $g_{\mu\nu}(x^\alpha;\epsilon)$, depending on the parameter
$\epsilon$, which satisfy the Einstein equations for all $\epsilon$,
and which, in the limit $\epsilon\rightarrow0$, become the
Schwarzschild metric functions, such as $g_{rr}=S^{-1}$. (Here
$S\equiv 1-2M/r$ and $M$ is the mass of the Schwarzschild spacetime;
we use units throughout in which $c=G=1$.)  NPS theory amounts to the
approximation
\begin{equation}\label{pertbasic}
g_{\mu\nu}(x^\alpha;\epsilon)\approx
g_{\mu\nu}(x^\alpha;\epsilon)|_{\epsilon=0}+
\epsilon\frac{\partial}{\partial\epsilon}
g_{\mu\nu}(x^\alpha;\epsilon)|_{\epsilon=0}\ .
\end{equation}

\subsection{Choice of expansion parameter}

It is of some practical importance to realize that the choice of the
expansion parameter can have a considerable effect on the range over
which perturbation theory gives a good approximation. Let us imagine
that we introduce a new parameter $\epsilon'$ which is a function of
$\epsilon$ such that $ d\epsilon'/d\epsilon$ approaches unity as
$\epsilon\rightarrow0$. If we take $\epsilon'$ to be the basis of our
perturbation approach, the approximation becomes
\begin{eqnarray}
g(x^\alpha;\epsilon)&=g(x^\alpha;\epsilon(\epsilon')) =
g_{\mu\nu}(x^\alpha;\epsilon(\epsilon'))|_{\epsilon'=0}+
\epsilon'\frac{\partial}{\partial\epsilon'}
g_{\mu\nu}(x^\alpha;\epsilon(\epsilon'))|_{\epsilon'=0}+{\cal
O}(\epsilon'^2)\nonumber\\ &=
g_{\mu\nu}(x^\alpha;\epsilon(\epsilon'))|_{\epsilon=0}+
\left[\epsilon\frac{\partial}{\partial\epsilon'}
g_{\mu\nu}(x^\alpha;\epsilon(\epsilon'))|_{\epsilon=0}\right]
\left\{\frac{\epsilon'}{\epsilon}\right\}+{\cal O}(\epsilon'^2)\
.\label{prime}
\end{eqnarray}
At $\epsilon=0$ the derivative of $g_{\mu\nu}$ with respect to
$\epsilon$ and with respect to $\epsilon'$ have the same values, so
for a given spacetime --- that is, for a given value of $\epsilon$ ---
the nonspherical perturbation in (\ref{prime}) differs from that in
(\ref{pertbasic}) by the factor
$\left\{\epsilon'/\epsilon\right\}$. Computed energies (which are
quadratic in the nonspherical perturbations) will differ by the square
of this ratio. Different choices of parameterization will change this
factor and affect the accuracy of the linearized approximation.

To show the effects of this parameterization dependence, we take as an
example Misner data \cite{price_pullin94}\cite{misner} for two holes.
The initial separation of the holes, in units of the mass of the
spacetime, is described by Misner's parameter $\mu_0$.  The metric
perturbations, however, are not analytic in $\mu_0$ as
$\mu_0\rightarrow0$, so $\mu_0$ cannot be used as the expansion
parameter in (\ref{pertbasic}).  The actual expansion parameter used
by Price and Pullin, was a function of $\mu_0$ denoted $\kappa_2$. We
consider here what would be the results of perturbation theory done
with the expansion parameter
\begin{equation}\label{eq.diffk}
\epsilon=\frac{\kappa_2}{1-k\kappa_2}\ .
\end{equation}

Figure 2 shows the results, along with the energies computed by
numerical relativity applied to full nonlinear evolution
\cite{anninos_etal93}. For all choices of $k$ the agreement between
perturbation theory and numerical relativity is good at sufficiently
small initial separation (sufficiently small $\mu_0$), but as $\mu_0$
grows larger, the agreement increasingly depends on the which
parameterization is used.  The $k=0$ parameterization, the parameter
of the Price-Pullin paper, is a reasonably good approximation even up
to separations ($\mu_0>1.36$) for which the initial apparent horizon
consists of two disjoint parts. For positive values of $k$ the
agreement is less impressive, while for $k=-4$, it appears that
perturbation theory is giving excellent answers for initial data that
are very nonspherical. Clearly the $k=-4$ parameterization is
``better,'' at least for the purpose of computing radiated
energy. There exist yet better choices; in principle a
parameterization could be found for which the energy computed by
linearized theory is perfect for any initial separation. The crucial
point is that we have no {\em a priori} way of choosing what is and
what is not a good parameterization. The choice of expansion parameter
$\kappa_2$ was made in the Price-Pullin analysis, because it occurred
naturally in the mathematical expressions for the initial
geometry. There was no {\em a priori} reason for believing it to be a
particularly good, or particularly bad parameterization. This point
will be discussed again, in connection with numerical results
presented in Sec.~IV.

The fact, demonstrated in Fig.~2, that the results of linear
perturbation theory are arbitrary may seem to suggest that
perturbation answers, from a formal expansion or numerical initial
data, are of little value.  It should be realized, however, that the
arbitrariness exhibited in Fig.~2 is simply a demonstration of the
fact that linearized perturbation results are uncertain to second
order in the expansion parameter. The fact that the results for
different parameterizations start to differ from each other around
$\mu_0\approx1.5$ simply signals that $\kappa_2$ is around unity. (In
fact, $\kappa_2\approx0.24$ for $\mu_0=1.2$.) Higher order uncertainty
is an unavoidable feature in the range where the expansion parameter
is of order unity. But there is a potential misunderstanding about the
meaning of ``expansion parameter around unity.'' To see this consider
a change to a new expansion parameter $\epsilon=10^{-4}*\kappa_2$. The
new expansion parameter $\epsilon$ is of order unity for
$\mu_0\approx7$, yet we know that perturbation fails dramatically for
such a large value of $\mu_0$. The issue here is that we need some way
of ascribing an appropriate ``normalization'' to the expansion
parameter. A sign that the normalization is good is that
physically-based measures of distortion start getting large for
$\epsilon$ around unity.  If we had reliable measures of this type
then we could have some confidence about the range of the the
expansion parameter for which we could neglect second order
uncertainty, whether due to parameter arbitrariness or the omission of
higher order terms in the calculation. One can formulate interesting
measures for the normalization of the expansion parameter, such as the
extent to which the linearized initial conditions violates the exact
Hamiltonian constraint \cite{suen}. Most such measures are useful only
for finding a very rough normalization for $\kappa_2$ (equivalently,
for roughly finding the range in which linearized perturbation theory
is reliable). The only reliable procedure for this is to carry out
computations of radiated waveforms and energy to second order in the
expansion parameter. The ratio of second order corrections to first
order results gives the only direct measure of the reliability of
perturbation results. If one computes an energy for which the second
order correction to the first order result is 10\%, then one knows
that the third order correction (due to a change in parameterization
or an inclusion of third order terms in the computation) will be on
the order of 1\%.

\subsection{Treating nonlinear initial data as a perturbation expansion}

We turn now to the central question of this paper: How does one apply
perturbation theory to numerically generated initial data?  To do this
we consider our numerical initial data to be initial data for a
solution in a parameterized family $g_{\mu\nu}(x^\alpha;\epsilon)$
corresponding to $\epsilon=\epsilon_{\rm num}$. The application of
perturbation theory is equivalent to replacing
$g_{\mu\nu}(x^\alpha;\epsilon_{\rm num})$ by
\begin{equation}\label{eq.linnum}
g_{\mu\nu}(x^\alpha;\epsilon)|_{\epsilon=0}+
\frac{\partial}{\partial\epsilon}
g_{\mu\nu}(x^\alpha;\epsilon)|_{\epsilon=0}\epsilon_{\rm num}.
\end{equation}
An added familiar complication is that we can introduce a family of
coordinate transformations $x^\alpha=x^\alpha(x^{\mu'};\epsilon)$
which reduces to $x^\alpha =x^{\alpha'}$ for
$\epsilon\rightarrow0$. Such a transformation takes the original
family to a new family $g^{'}_{\mu\nu}(x^{\alpha'};\epsilon)$, which
satisfies the same requirements as the original family. We follow
Moncrief\cite{moncrief74} in constructing, from the 3-metric
$\gamma_{ij}$ on constant-$t$ surfaces, quantities $q_i$, which are
invariant to first order in $\epsilon$ (``gauge invariant''), for
coordinate transformations. The construction of these Moncrief $q_i$
is done in two steps. First, the multipole moments of the metric are
extracted. In practice this is done by multiplying the metric
functions by certain angular factors and integrating over
angles. Since we are only interested in quadrupole and higher order
for radiation, this step also eliminates the spherically symmetric
background parts of the metric function. The second step is to form
linear combinations of these multipoles and of their derivatives with
respect to radius. We symbolically represent the process of forming
these quantities as
\begin{equation}\label{monconstr}
q_i=Q_i(\gamma_{ij}, \partial_r \gamma_{ij})\ .
\end{equation}
Here the symbol ``$Q_i$'' represents the process of multiplying by
angular functions and integrating, then multiplying by certain
functions of $r$ and taking linear combinations of the results. (Our
notation here disagrees with that of Moncrief\cite{moncrief74} in a
potentially confusing way. Moncrief's perturbation quantities are
independent of the size of $\epsilon$. In order to have definitions
that can be applied to numerical data we use quantities that -- to
first order -- are proportional to $\epsilon$.)

The Moncrief gauge invariants play two different roles. For even
parity one of the gauge invariants, $q_2$, is a constraint; it
vanishes in linearized theory as a result of the initial value
equations. In linearized theory, the remaining Moncrief quantities,
denoted $q_1$ here, satisfy wave equations $L(q_1)=0$, the
Regge-Wheeler equation in odd parity and Zerilli equation in even
parity.

{}From our numerical data we construct the quantities $q_i$ precisely
according to (\ref{monconstr}).  Our numerically constructed
``perturbation'' quantities will not be invariant under coordinate
transformations, but rather will transform as $q'_i=q_i+{\cal
O}(\epsilon_{\rm num}^2)$. Similarly, the linearized constraint, $q_2$
will not vanish, but will be of order $\epsilon_{\rm num}^2$. The
numerically constructed wavefunctions $q_1$ will satisfy $L(q_1)={\cal
O}((\epsilon_{\rm num})^2)$, where $L$ is the Regge-Wheeler or Zerilli
wave operators.

The use of NPS methods is equivalent to ignoring the second order
terms in the wave equations.  The wavefunction $q_i$ can then be
propagated from the initial hypersurface forward and the radiation
waveforms extracted from it. To evolve $q_1$ off the initial
hypersurface, however, requires the initial time derivative $\partial
q_1/\partial t$. This can be computed from the initial extrinsic
curvature, but some care is needed. Indeed, the possible ambiguities
that arise here are the justification for the somewhat protracted
discussion in this section.

If ${\bf n}$ is the future-directed unit normal to the initial hypersurface
then the rate at which the 3-metric is changing is given by
\begin{equation}\label{K=Lng}
K_{ij}=-\frac{1}{2}{\cal L}_{\bf n}\,\gamma_{ij}\ ,
\end{equation}
where $K_{ij}$ is the extrinsic curvature and ${\cal L}_{\bf n}$ is
the Lie derivative along the unit normal.  The unit normal is related
to the derivative with respect to Schwarzschild time by
$\partial/\partial t=S^{1/2}\ {\bf n}$.  The time derivative of the
Moncrief function then can be written
\begin{displaymath}
\partial q_1/\partial t=S^{1/2} {\cal L}_{\bf n}q_1
\end{displaymath}\begin{equation}
=S^{1/2}{\cal L}_{\bf n}Q_1(\gamma_{ij},
\partial\gamma_{ij}/\partial r)\ .
\end{equation}

To evaluate the right hand side we need to know how $Q_1$ changes when
it is Lie dragged by ${\bf n}$. Since $Q_1$ depends only on
$\gamma_{ij}$ it might appear that one need only Lie drag
$\gamma_{ij}$ to find the change in $Q_1$, and that ${\cal L}_{\bf
n}Q_1=Q_1({\cal L}_{\bf n}\gamma_{ij},\partial{\cal L}_{\bf
n}\gamma_{ij}/\partial r)$. From this it would follow that $\partial
q_1/\partial t=-2S^{1/2} Q_1(K_{ij}, \partial K_{ij}/\partial r)$. It
is important to note that this is {\em not} the correct relationship
between $K_{ij}$ and the cauchy data for the wave equation. The
fallacy in this procedure lies in the fact that $q_1$ must be computed
from the 3-metric on a slice for which Schwarzschild time is constant
(to first order in $\epsilon_{\rm num}$). Lie dragging by ${\bf n}$
moves the 3-metric to a surface that is not (to first order) a
constant time surface. The cure is clearly to compare quantities on
surfaces of constant $t$ by using ${\cal L}_t\equiv S^{1/2}{\cal
L}_{\bf n}$.  It is the Schwarzschild time derivative that commutes
with the Schwarzschild radial derivative ${\cal L} _t (\partial
/\partial r)^a=0$.  The correct prescription then follows from
\begin{displaymath}
\partial q_1/\partial t=S^{1/2}{\cal L}_{\bf n}q_1
\end{displaymath}\begin{displaymath}
=Q_1(S^{1/2}{\cal L}_{\bf n}\gamma_{ij},
\partial (S^{1/2}{\cal L}_{\bf n}\gamma_{ij})/\partial r)
\end{displaymath}
\begin{equation}\label{qdot}
= -2 Q_1(S^{1/2} K_{ij},
\partial (S^{1/2}K_{ij})/\partial r)\ .
\end{equation}

We note that the perturbed Schwarzschild metric does have a shift
vector $\beta_i$ of order $\epsilon$, and in principle the shift
vector influences the time development of $\gamma_{ij}$ according to
$\partial_{t} \gamma_{ij} = \partial_{t'} \gamma_{ij} + 2
\nabla_{(i}\beta_{j)}$, where $t'$ is a time coordinate in which the
shift vector vanishes.  But the shift vector can be considered to be
``pure gauge.''  It is necessary if one wants a complete specification
of the coordinates and the metric components, but its value is a
matter of choice, and is not necessary for a complete specification of
the physics.  The initial value, and evolution, of the gauge invariant
quantity $q_1$ is invariant with respect to the choice of $\beta_i$,
and $q_1$ carries all the (physically meaningful) information about
gravitational waves.

The evaluation of $q_1$ from (\ref{monconstr}) and $\partial
q_1/\partial t$ from (\ref{qdot}) completes the extraction, from the
numerical data for $\gamma_{ij}, K_{ij}$ of the cauchy data for the
Regge-Wheeler or Zerilli wave equation.  An alternative procedure
arises if one uses the scalar wave-equations derived from the
perturbative reduction of the nonlinear wave-equation for the
extrinsic curvature which arises in a new explicitly hyperbolic form
of the Einstein equations\cite{aacby95}.  In this system, the scalar
wave equations are one order lower in time derivative from the usual
Regge-Wheeler and Zerilli equations, so the Cauchy data consists of
the extrinsic curvature and its time-derivative (which involves the
3-dimensional Ricci curvature).

{}From the above it is clear that linearized evolution should give good
accuracy when applied to numerically generated initial data with
sufficiently small deviations from sphericity. For initial data which
are known in analytic form one can, of course, apply linearized theory
even to cases in which initial deviations from sphericity are only
marginally small.  The results in Fig.~2, for example, show that the
results of such application of perturbation theory give reasonable
accuracy for values of $\mu_0$ at which an initial horizon is highly
distorted. It is worrisome to apply linearized evolution to marginally
nonspherical initial data, which do not, for example, satisfy the
constraint $q_2=0$ with reasonable accuracy. Such a procedure ---
linear evolution of nonlinear initial data --- has, among other
disadvantages, no clear theoretical framework.

\subsection{Calculating radiated energy by ``forced linearization''}

We wish to point out here that NPS methods can be used more broadly,
and a procedure we call ``forced linearization'' can be applied to
numerically generated initial data in a way that amounts to extracting
the linearized part of the data and evolving linearly.  This procedure
circumvents the difficulty of performing formal linearization to data
which is only known numerically.  We imagine that we start with an
initial value problem in which there is some adjustable parameter,
call it $\mu$, such that $\mu=0$ corresponds to the Schwarzschild
initial data.  There is no requirement that the family of solutions
$g_{\mu\nu}(x^\alpha;\mu)$ be analytic in $\mu$ at $\mu\rightarrow0$.
There may be additional parameters, call them $p_i$, such as the
parameters governing the initial momenta of holes. To apply forced
linearization we fix the values of the $p_i$ and make a choice of
$\mu$ such that the computed initial data $\gamma^{\rm
vns}_{ij},K^{\rm vns}_{ij}$ are ``very nearly spherical.'' One
criterion for this would be that $q_2$ is very small. We then
interpret this initial data as being essentially linearized data, to
which the approximation in (\ref{eq.linnum}) applies. We extract
multipoles, form a gauge invariant wave function $q_1$, and evolve it
with the Zerilli or Regge-Wheeler equation, all as described
above. The result of this will be a late-time waveform $q_1^{\rm
vns}(r,t)$ and the energy $E^{\rm vns}$ that it carries.  The next
step is to characterize the results with a well behaved gauge
invariant parameter. To do this we choose some fiducial radius $r_{\rm
fid}$, and evaluate $\epsilon^{\rm vns}\equiv q_1(r_{\rm fid},t=0)$
the gauge invariant wave function of the initial hypersurface at this
radius.

Next, we leave the $p_i$ unchanged, but choose a larger value of $\mu$
for which the numerically generated initial data set $\gamma^{\rm
mrgnl}_{ij},K^{\rm mrgnl}_{ij}$ is ``marginal'' in that it corresponds
to deviations from sphericity large enough so that it differs
significantly form linearized initial conditions; one sign of this
would be that the condition $q_2=0$ is significantly violated. For
this data set we go through the same procedure as above in
characterizing the data set by a parameter $\epsilon^{\rm mrgnl}\equiv
q_1(r_{\rm fid},t=0)$. For this marginally spherical initial data we
take the solution for the wavefunction and energy to be
\begin{equation}
q_1^{\rm mrgnl}(r,t)=\left( \frac{\epsilon^{\rm mrgnl}}{\epsilon^{\rm
vns}}\right)q_1^{\rm vns}(r,t)
\ \ \ \ \
E^{\rm mrgnl}=\left( \frac{\epsilon^{\rm mrgnl}}{\epsilon^{\rm
vns}}\right)^2E^{\rm vns} \ .
\end{equation}
The idea underlying this method is that the very nearly spherical data
give us the solution for for $\partial
g_{\mu\nu}/\partial\epsilon|_{\epsilon=0}$.  For the marginal initial
data set we then need only multiply this initial data by the
appropriate factor telling us how much larger is the linear part of
the nonsphericity than that of the very nearly spherical initial
data. The success of forced linearization requires then that
$\epsilon$ evaluated at $r_{\rm fid}$ be a well behaved
parameterization of the linearized part of the nonsphericity in the
numerical data. Since our expansion parameter $\epsilon$ is the
magnitude of the perturbation, it will be a good expansion parameter
as long as it is evaluated in a region where the nonlinear deviations
from sphericity are small, i.e., where (\ref{eq.linnum}) is a good
approximation. For this reason it is important that $r_{\rm fid}$ be
chosen fairly large. For processes of the type pictured in Fig.~1, the
deviations from sphericity fall off quickly in radius, so that at
large enough $r$ one can be certain that the initial data are an
excellent approximation to linearized data. Evidence for this is that
the violations of the $q_2=0$ constraint are always confined to small
radii. One easily implemented check on the forced linearization
procedure is to look at the factor $\epsilon^{\rm mrgnl}/\epsilon^{\rm
vns}$ and confirm that it is independent of $r$ for $r>r_{\rm
fid}$. In Section III we show that this test is easily passed by a
numerical example, and that the results of forced linearization are
essentially the same as those of formal linearized theory.

\section{Extraction of perturbations from numerical data}

Here we assume that the reader has numerical solutions for the
3-metric on an approximately t=const surface.  The first step in
applying NPS to numerical results is to transform to coordinates which
are ``almost Schwarzschild'' coordinates. It is assumed that the
numerical $\gamma_{ij}$ and $K_{ij}$ are expressed in a coordinate
system $R,\theta,\phi$ in which the approximate spherical symmetry is
manifest. This means that $K_{ij}$ and ratios like
\begin{equation}
\frac{
\gamma_{R\theta}
}
{\sqrt{\gamma_{\theta\theta}}}
\ \ \ \ \ \
\frac{
\gamma_{R\phi}
}
{\sqrt{\gamma_{\theta\theta}}}
\ \ \ \ \
\frac{
\gamma_{\theta\phi}
}
{\gamma_{\theta\theta}}
\end{equation}
must be small. They all are, in fact, formally of order
$\epsilon_{\rm num}$, so if they are not all reasonably
small compared to unity there is little reason to think that
NPS will work. A Schwarzchild-like areal radial coordinate $r$ needs to be
introduced. This can be defined as a function of $R$ by
\begin{equation}
r\equiv\left( \int\,\gamma_{\theta\theta}\,\gamma_{\phi\phi}\,d\Omega\
\right)^{1/4}/4\pi.
\end{equation}
where the integral is taken on a surface of constant $R$. The metric
component $\gamma_{rr}$, in terms of this quantity, gives us another test
of how close the geometry is to that of a constant time Schwarzschild
slice. The quantity
\begin{displaymath}
r\left(1-1/\gamma_{rr}\right)
\end{displaymath}
should be nearly equal to the constant $2M$, where $M$ is the mass of
the spacetime. The variability of this quantity in $r, \theta$, and $\phi$,
is formally of order $\epsilon_{\rm num}$.

There are, of course, other ways of specifying the Schwarzschild-like
coordinates. We could, for example, have defined
$r^2\equiv\gamma_{\theta\theta}$ All these coordinate choices,
however, should agree to order $\epsilon_{\rm num}$ and are therefore
equivalent within a linearized gauge transformation.

To compute the gauge invariant perturbation functions, we first assume
that an $\ell m$ multipole of the 3-metric may be expanded as
\begin{equation}\label{mncrform}
\gamma_{ij}=c_1 (\hat e_1)_{ij} +c_2 (\hat e_2)_{ij}
+h_1 (\hat f_1)_{ij} + {H_2 \over S} (\hat f_2)_{ij}+
r^2 K (\hat f_3)_{ij}+ +r^2 G (\hat f_4)_{ij}
\end{equation}
where, for clarity, we have suppressed multipole indices and have
replaced Moncrief's $h_1$ and $h_2$ odd parity perturbation functions
with $c_1$, $c_2$.  The multipole moments $c_1, c_2, h_1, H_2, K,$ and
$G$ are computed by projection onto the relevant spherical harmonics
which can be found in Moncrief\cite{moncrief74}.  Explicit formulas
for the important special case of even parity, axisymmetric
perturbations may be found in Ref.~\cite{abhss}.

For odd parity perturbations, one function can be constructed from the
amplitudes $c_1$ and $c_2$ which is gauge invariant
and satisfies the Regge-Wheeler equation (below),
\begin{equation}
Q^{\times}_{\ell m} = \sqrt{2\frac{(\ell+2)!}{(\ell-2)!}}\,
\left[c_1+{1 \over 2} \left ( {\partial c_2 \over \partial r}
- {2 \over r} c_2 \right ) \right ] {S \over r}.
\end{equation}

The situation for even parity perturbations is more
complicated.  Two gauge invariant functions may be formed out of
the multipole moments:
\begin{eqnarray}
k_1&=& K + {S \over r} (r^2 \partial_r G- 2 h_1)
\\
k_2 &=&{1 \over 2S} \left[ H_2 - r \partial_r k_{1}
- \left(1-\frac{M}{rS}\right)k_1+ S^{1/2}\partial_r
(r^2 S^{1/2} \partial_r G - 2 S^{1/2} h_1) \right]
\end{eqnarray}
{}From $k_1$ and $k_2$ it is possible to form two new functions,
one which is radiative and one which is
equivalent to the perturbed hamiltonian constraint
\begin{eqnarray}
q_1 &=& 4 r S^2 k_2 + \ell (\ell+1) r k_1
\\
q_2 &=& \partial_r [ 4 r S^2 k_2 + \ell (\ell+1) r k_1 ]
+ \ell(\ell+1) [2 S k_2 + (1-M/\{rS\}) k_1 ].
\end{eqnarray}
The scaled function
\begin{equation}
Q^+_{\ell m}={ q_1 \over \Lambda}\sqrt{2(\ell-1)(\ell+2)
\over \ell(\ell+1)}\ ,
\end{equation}
with
\begin{displaymath}
\Lambda\equiv(\ell-1)(\ell+2)+6M/r\ ,
\end{displaymath}
satisfies the Zerilli equation (below).

The time derivatives of the radiative gauge invariant
functions $Q^{\times}_{\ell m}$ and $Q^+_{\ell m}$
are found by substituting $\sqrt{1-2M/r}K_{ij}$ for $\gamma_{ij}$
in the multipole moment computation and forming the
same combinations of moments.

The wavefunctions $Q^{\times}_{\ell m}$ and $Q^{+}_{\ell m}$
obey the Regge-Wheeler and Zerilli wave equations respectively:
\begin{eqnarray}
{\bf L} Q^{\times}_{\ell m} +V^{\times}_{\ell} Q^{\times}_{\ell m} &=& 0 \\
{\bf L} Q^{+}_{\ell m} +V^{+}_{\ell} Q^{+}_{\ell m} &=& 0
\label{eq:zer}
\end{eqnarray}
where the wave operator appropriate to Schwarzschild
spacetime is
\begin{equation}
{\bf L} = {\partial ^2  \over \partial t^2} - {\partial ^2 \over \partial
r_*^2}
\end{equation}
in terms of the ``tortoise coordinate'' $r_*=r+2M \ln (r/2M-1)$,
and where the potentials are given by
\begin{equation}
V^{\times}_{\ell} = (1-2M/r) \left [ {\ell(\ell +1) \over r^2}
-{6M \over r^3} \right ]
\end{equation}
and,
\begin{equation}
V^{+}_{\ell}(r) = (1-2M/r) \left[{1\over\Lambda^2} \left (
{72M^2\over r^5}-{12M\over r^3}(\ell-1)(\ell+2)(1-3M/r)\right)
+{\ell(\ell-1)(\ell+1)(\ell+2) \over
r^2 \Lambda } \right].
\end{equation}

Once the Zerilli and Regge-Wheeler equations are integrated for
all the desired $\ell$ and $m$ modes, the total radiated energy
can be calculated from the asymptotic timeseries for
$Q_{\ell m}^+$ and $Q_{\ell m}^{\times}$:
\begin{equation}
{d E \over dt} = {1 \over 32 \pi} \sum_{\ell=2}^\infty
\sum_{m=-\ell}^\ell \left( {d Q_{\ell m}^+ \over dt}^2+
{d Q_{\ell m}^{\times} \over dt}^2 \right)
{}.
\label{eq:power}
\end{equation}

\section{Example of perturbation extraction}

In this section we demonstrate the extraction of a
perturbation from a numerical solution to the nonlinear constraint
equations -- the Misner data representing two black holes
at a moment of time symmetry.   The Misner 3-geometry may
be written \cite{price_pullin94} as
\begin{equation}
\label{eq:mismet}
dl^2 = \Phi(r, \theta, \mu_0)^4(S^{-1} dr^2 + r^2 d\Omega^2)
{}.
\end{equation}

The conformal factor $\Phi$ is given
by
\begin{equation}
\label{eq:miscon}
\Phi(r, \theta; \mu_0)=
1 + 2 (1+M/2R)^{-1}\sum_{l=2,4,...}^{\infty} \kappa_\ell \left({M \over R}
\right)^{\ell+1} P_\ell(\cos \theta),
\end{equation}
where
\begin{displaymath}
R\equiv(\sqrt{r}+\sqrt{r-2M}\,)^2/4
\end{displaymath}
and
\begin{equation}
\label{eq:miskap}
\kappa_{\ell}\equiv \left({1 \over 4 \sum_{n=1}^\infty (\sinh{n
\mu_0})^{-1}} \right)^{\ell+1}\, \sum_{n=1}^\infty {(\coth{n
\mu_0})^\ell \over \sinh n \mu_0 }\ .
\end{equation}

For this exercise, we pretend that the initial geometry is known only
numerically, so no explicit formal linearization can be done. The odd
parity perturbations vanish in the Misner solution. We compute the
even parity gauge invariant wavefunction for $\ell=2$ using numerical
evaluations of (\ref{eq:miscon}) - (\ref{eq:miskap}).  Specifically, we
compute $K$ and $H_2$ of (\ref{mncrform}) from
\begin{equation}
K= H_2 = \int d \Omega \Phi^4 Y_{20}\ .
\end{equation}
All the other moments in (\ref{mncrform}) vanish for the conformally
Schwarzschild metric of (\ref{eq:mismet}).  The function $Q_{20}^+$ is
evaluated at values of $r$ corresponding to the range $r_* = -20M$ to
$r_*=50M$.  The initial value of $Q_{20}^+$ (along with its
time-derivative which is zero for the Misner time-symmetric initial data)
provides initial values for integration of (\ref{eq:zer}).  At large
radius, $r= 100M$, the value of $\partial Q_{20}^+/\partial t$ is used in
(\ref{eq:power}) to compute the radiated energy.

First, in Fig.~\ref{fig.misnl} we show the result of directly
computing the gauge invariant function $Q_{20}^+$ from the nonlinear
initial data, integrating the Zerilli equation, and computing the
radiated energy.
For small values of $\mu_0$ the agreement with the explicitly
linearized data of Ref.~\cite{price_pullin94} is excellent.  At about
$\mu_0 \simeq 1.2$ the agreement breaks down and the qualitative
behavior becomes dramatically different.  It is interesting to note
that the apparent horizon encompassing both black holes does not exist
for $\mu_0 >1.36$, close to the dramatic reversal in the energy curve.

In Fig.~\ref{fig.miscon} the violation of the linearized constraint by
the nonlinear data is shown as a function of radius.  We plot the
ratio of the constrained gauge invariant function, $q_1$ to the
radiative function $q_2$ scaled in such a way as to compensate for
large violation at $r=2M$. The value of $q_2$ clearly grows much
faster than the radiative variable $q_1$ as the separation is
increased.

As discussed in Sec.~II, it is possible to obtain the results of
formal perturbation theory directly from the numerical data without
ever making reference to the analytic solution.  In
Fig.~\ref{fig.misfl} we demonstrate the application of the forced
linearization procedure to the nonlinear Misner data for various
values of the fiducial radius $r_{\rm fid}$. For very small values of
$\mu_0$, such as $\mu_0=0.5$, the geometry outside the event horizon
is everywhere well approximated by (\ref{eq.linnum}) and forced
linearization works even for small values of $r_{\rm fid}/M$. When
$\mu_0$ is larger than around $1.5$, on the other hand, the initial
geometry near the horizon contains significant nonlinear effects, and
large values of $r_{\rm fid}/M$ must be used to get results equivalent
to those of formal linearized theory.

As $r_{\rm fid}$ gets large, the results become indistinguishable from
those of formal perturbation theory reported in
Ref.~\cite{price_pullin94}.  For $r_{\rm fid}=30M$ the difference in
radiated energy for $\mu_0=3.0$ is less than $10^{-3} \%$.  This
high-accuracy equivalence deserves some explanation.  In particular,
why is forced linearization equivalent to formal linearization with
expansion parameter $\kappa_2$? Why is that expansion parameter
singled out? The equivalence is a result of two features of the way in
which the linearizations were done: First, both the formal
linearization of Ref.~\cite{price_pullin94}, and the forced
linearization results in Fig.~\ref{fig.misfl}, use precisely the same
coordinates. (The forced linearization results, in fact, are not based
on initial values that were generated by genuinely numerical
means. Rather, the closed form solutions for the Misner metric
functions were used. The ``almost-Schwarzschild'' coordinates of the
forced linearization, were precisely the same as the
``almost-Schwarzschild'' coordinates in
Ref.~\cite{price_pullin94}). Secondly, in the ``almost-Schwarzschild''
coordinate system, the parameter $\kappa_2$ is, to all perturbation
orders, the coefficient of the dominant nonsphericity at large radius.
Forced linearization (in the limit of large $r_{\rm fid}$) results in
a parameterization based on a gauge invariant measure of nonsphericity
at large radius. It therefore must be proportional to $\kappa_2$ and
produce results equivalent to those of the formal linearization of
Ref.~\cite{price_pullin94}, in which $\kappa_2$ was the expansion
parameter.

It should be understood that this does not imply that the parameter
$\kappa_2$ is {\em physically} singled out.  A first order change in
the ``almost-Schwarzschild'' coordinates will change the coefficient
of the dominant large-radius nonsphericity. We might, for example,
transform from the ``almost-Schwarzschild'' radial coordinate $r$ of
(\ref{eq:mismet}) to a new coordinate $r' \equiv r [1 + \kappa_{2}
P_2 (\cos \theta)]$. In this case the coefficient of the leading large
$r'$ term in the metric will be $\kappa_2+{\cal O}(\kappa_2^2)$, and
the results of forced linearization with the resulting ``numerical''
data will differ, when perturbations are large, from the results in
Ref.~\cite{price_pullin94}. The forced linearization will have induced
an expansion parameter different from $\kappa_2$.

AMA was supported by National Science Foundation
grant PHY 93-18152/ASC 93-18152 (ARPA supplemented).
RHP was supported by the National Science Foundation under grants
PHY9207225 and PHY9507719.

\begin{figure}
\caption{ Spacetime regions for coalescence. The ``legs of the
trousers'' represent the world tubes of two compact objects before
coalescence; region I cannot be considered to be nearly spherical.
The objects coalesce in region II which is also highly nonspherical,
but lies inside a horizon.  Region III, above the hypersurface and
outside the horizon, can be treated as a nearly spherical spacetime. }
\end{figure}

\begin{figure}
\caption{The effect of a change of expansion parameter. Results are
given for the energy radiated, as a function of $\mu_0$, during the
head-on collision of two black holes (Misner initial data).  The
results of numerical relativity are compared with linearized theory
for different choices of expansion parameters.}
\end{figure}

\begin{figure}
\caption{Radiated energies from nonlinear Misner data.  Energies
computed by integration of the Zerilli equation are compared for
initial perturbations calculated by explicit linearization of the
Misner data (solid line) and initial perturbations extracted directly
from the nonlinear Misner data (dashed line).  }\label{fig.misnl}
\end{figure}

\begin{figure}
\caption{Violation of the linearized hamiltonian constraint.
The ratio of gauge invariant functions $q_1/q_2$ scaled by
the factor $r-2M$ is plotted as a function of tortoise coordinate
$r_*$.  Curves are shown for $\mu_0 = 0.2, 0.4, 0.6, 0.8, 1.0, 1.2, 1.4$.
The largest constraint violation occurs for $\mu_0=1.4$.
}\label{fig.miscon}
\end{figure}

\begin{figure}
\caption{Radiated energies from forced linearization procedure.
Radiated energy is plotted as a function of Misner separation
parameter $\mu_0$ for various values of $r_{\rm fid}$.  The curve for
$r_{\rm fid}=30M$ is indistinguishable from the formal perturbation
theory result of Price and Pullin.  }
\label{fig.misfl}
\end{figure}
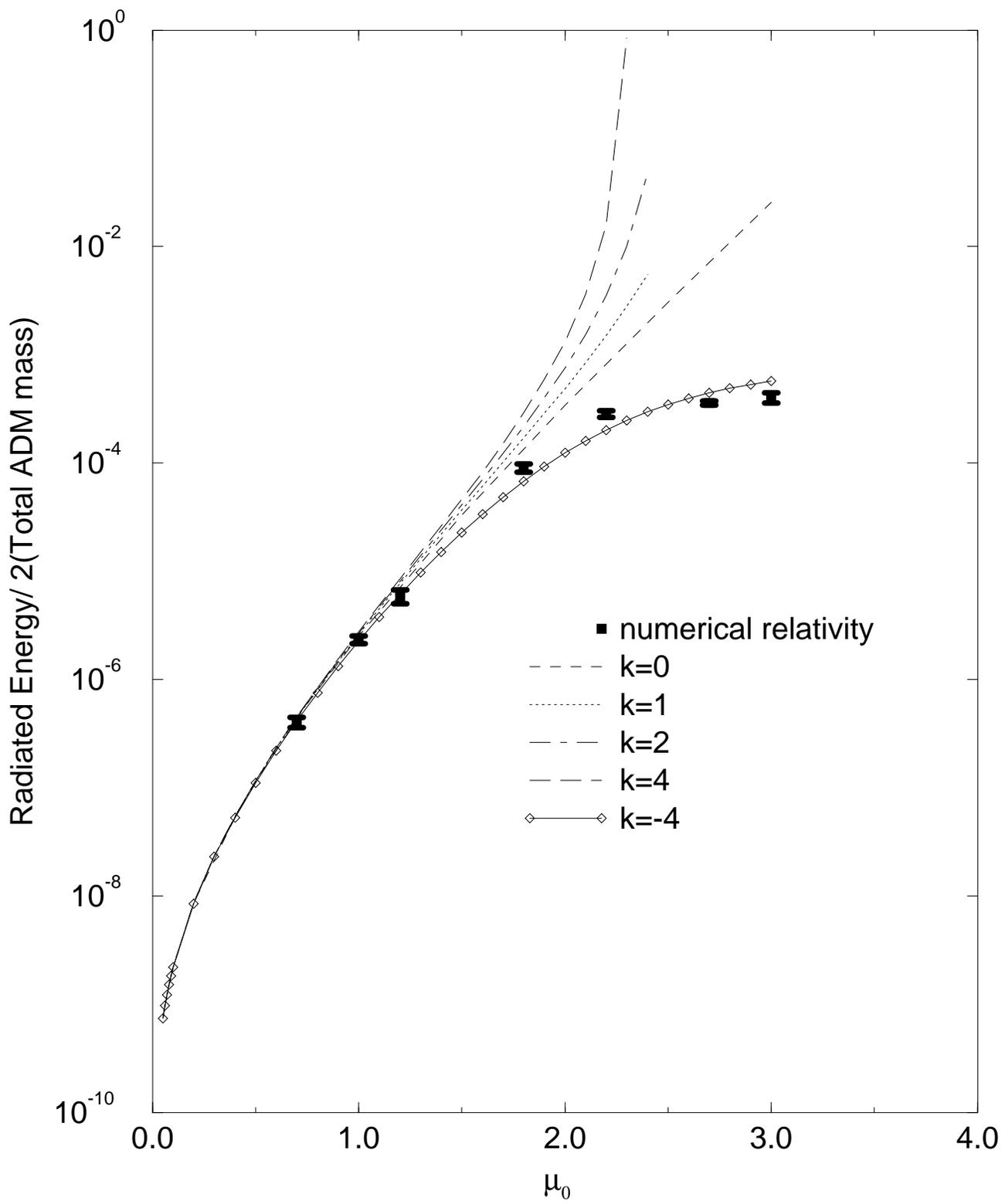
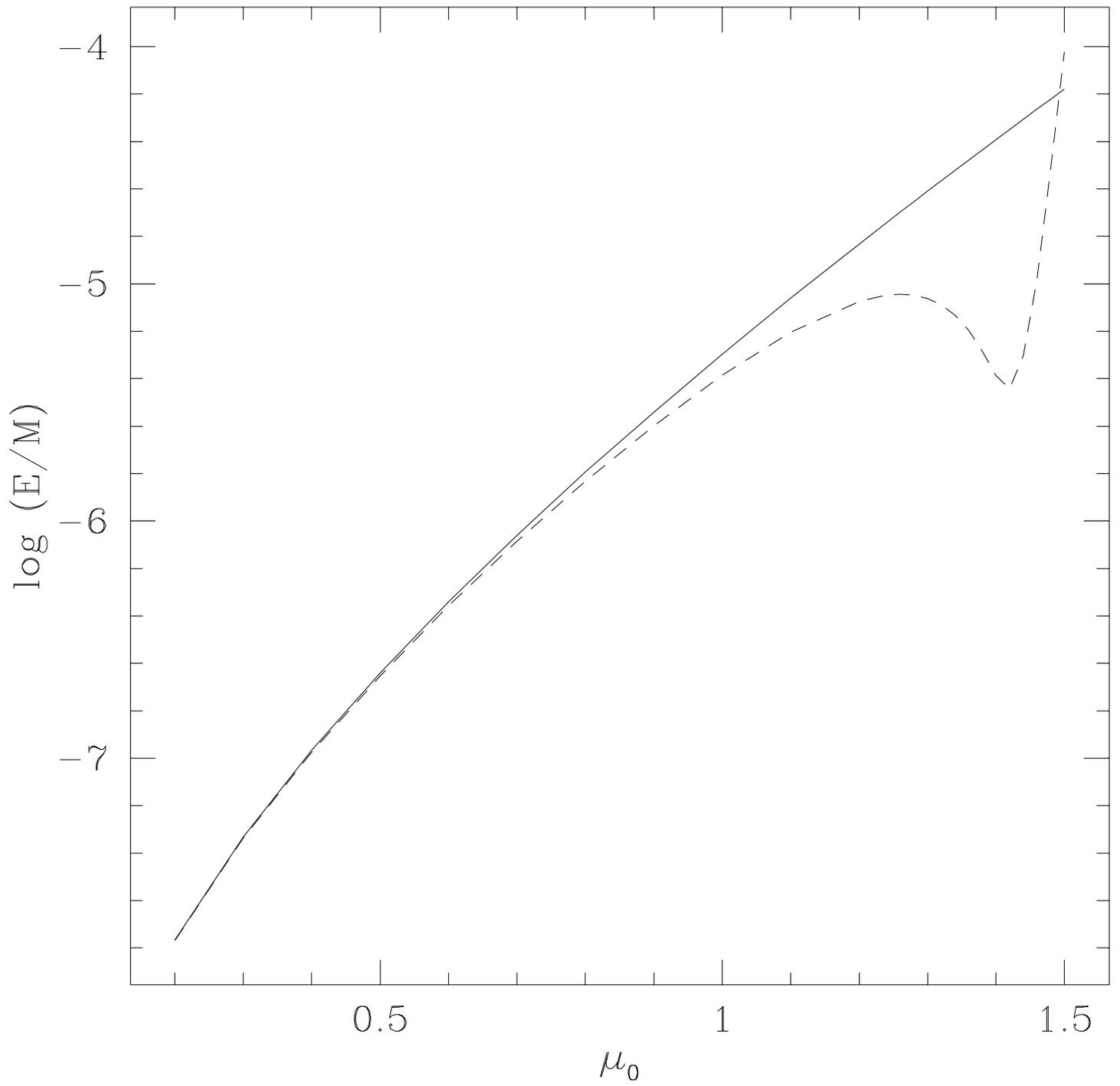
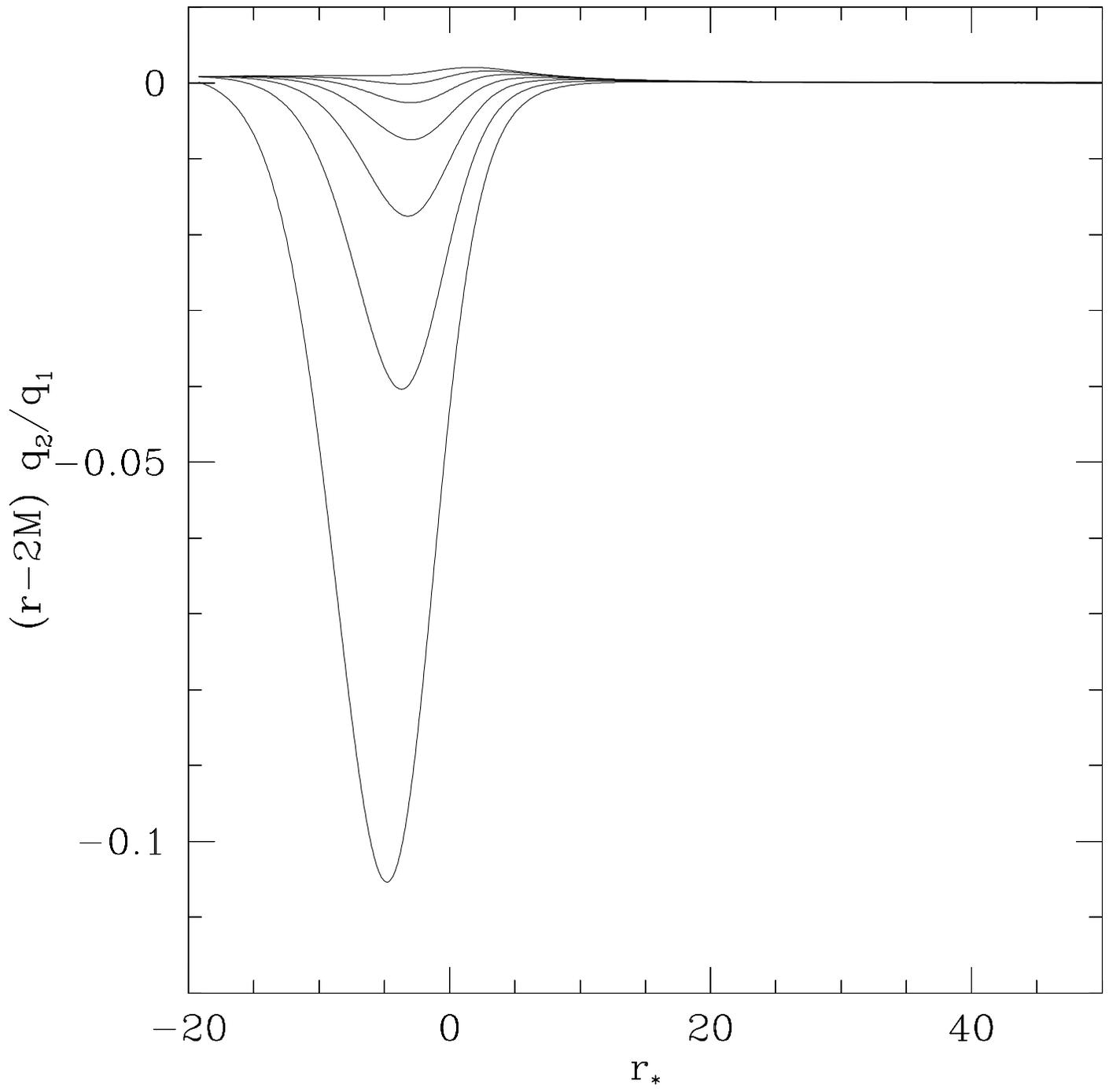
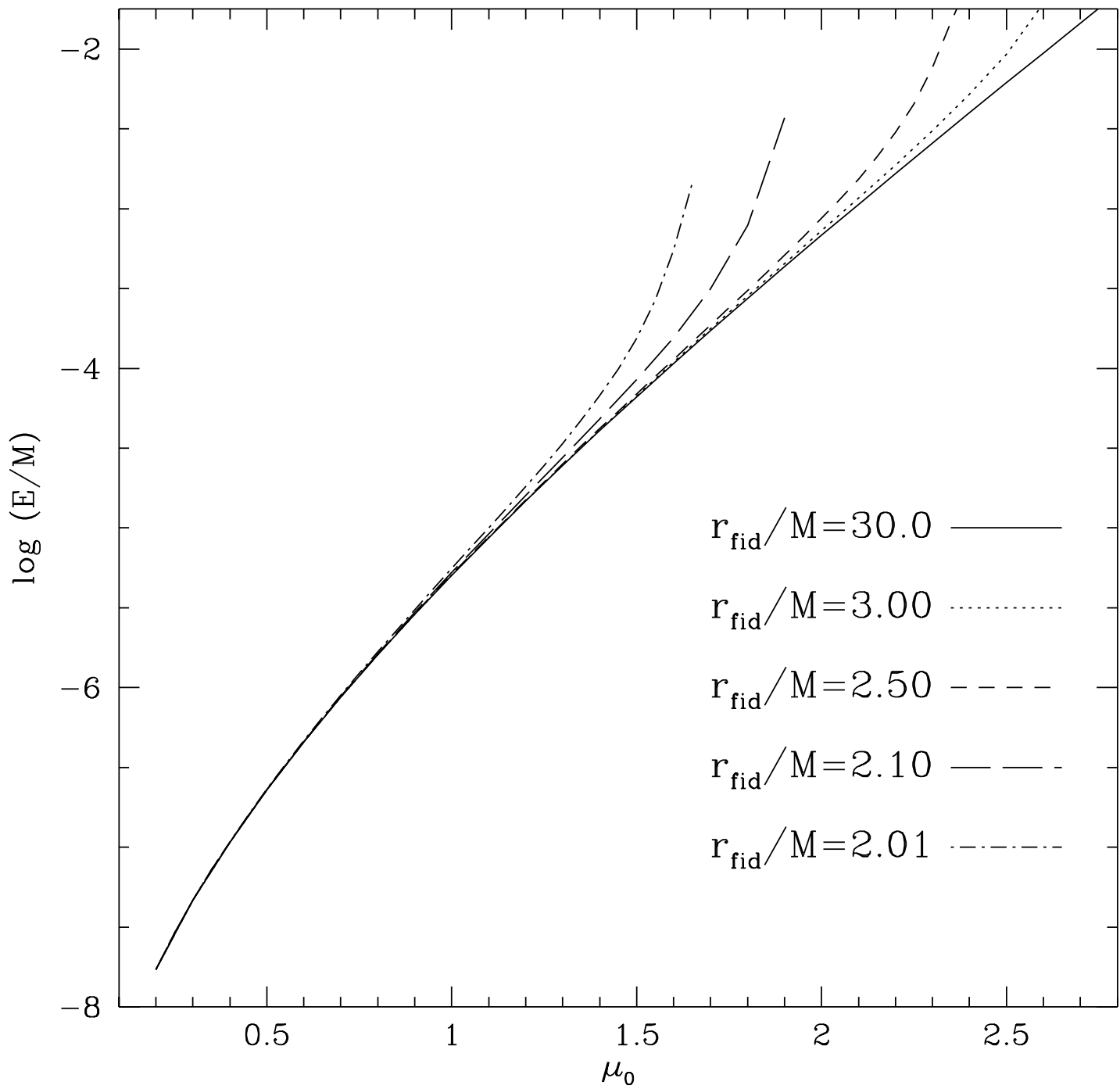
\end{document}